# Do Streetscapes Still Matter for Customer Ratings of Eating and Drinking Establishments in Car-Dependent Cities?

**Abstract**

This study examines how indoor and outdoor aesthetics, streetscapes, and neighborhood features shape customer satisfaction at eating and dining establishments (EDEs) across different urban contexts, varying in car dependency, in Washington, DC. Using review photos and street view images, computer vision models quantified perceived safety and visual appeal. Ordinal logistic regression analyzed their effects on Yelp ratings. Findings reveal that both indoor and outdoor environments significantly impact EDE ratings, while streetscape quality's influence diminishes in car-dependent areas. The study highlights the need for context-sensitive planning that integrates indoor and outdoor factors to enhance customer experiences in diverse settings.

**Highlights**

- WalkScore positively correlates with customer satisfaction, increasing the likelihood of a place having a higher rating category by 3% per point.
- Perceived Safety of surrounding environment has a positive and statistically significant relationship with rating score, with a 0.1-point increase raising the likelihood of rating score being in a higher category by 8.7 times.
- Higher car dependency of an EDE moderates the positive effect of perceived safety on rating score.

**Authors names and affiliations**

Chaeyeon Han[a], Seung Jae Lieu[a], Uijeong Hwang[b], Subhrajit Guhathakurta[a]

[a]School of City and Regional Planning, Georgia Institute of Technology, 245 4th Street NW, Atlanta, GA, 30332, United States
[b]Atlanta Regional Commission, 229 Peachtree St. #100, Atlanta, GA, 30303, United States

## 1. Introduction

In recent years, digital place rating platforms like Yelp and Google Maps have transformed how people choose eating and drinking establishments like bars, cafes, and restaurants, henceforth called EDEs, by offering more detailed information than word-of-mouth recommendations (Xiang et al., 2015; Book et al. 2018). These platforms provide not only operational details such as parking availability and business hours but also user-generated photos (UGPs), text reviews, and rating scores, significantly influencing customers' expectations about their dining experience in the EDEs (Ma et al., 2018; Li et al., 2023). Notably, the photo reviews give a visual impression to prospective visitors about the atmosphere of the indoor space, including factors such as furniture, lighting, and décor. The power of these platforms extends beyond information-sharing, as they shape perceptions of the quality of a place itself, driving people to visit based on digital reviews and aesthetic presentations, even in the absence of personal referrals.

In addition to the quality of food, service, and ambiance of the EDEs, the attractiveness of the surrounding built environment, especially that of walkability and the aesthetic quality of the streetscape have been recognized as key factors that increase people's satisfaction to points-of-interests (POIs) (Li et al., 2018; Ki & Lee, 2021; Koo et al., 2021, 2023). These metrics quantify aspects of urban space that influence how inviting, beautiful, safe, and functional a location appears to potential visitors. In addition, well-connected and more accessible streets within denser networks in mixed-use areas are recognized as a draw for local businesses, as they offer safer and more convenient routes for customers (Pivo & Fisher, 2011).

The concept of "servicescape," rooted in business and marketing fields, provides a useful lens for examining how the attractiveness of a location impact customer satisfaction with the place (Bitner, 1992). Initially focused on indoor factors (Ryu & Jang, 2008; Ryu & Han, 2010; Horng & Hsu, 2021), servicescape theory has expanded to include the exterior facades and the streets surrounding businesses (Koo et al., 2023; Wolf, 2003, 2004; Yüksel, 2013). This holistic perspective underscores the interconnectedness of indoor and outdoor environments in shaping customer experiences.

Understanding which factors of a servicescape makes a place satisfying is critical for supporting a reciprocal relationship between attractive local businesses and high-quality urban environments (Koo et al., 2023). Attractive places not only draw visitors but also enhance a city's walkability, creating a feedback loop that drives economic vitality while delivering social, environmental, and health benefits (Koo et al., 2023; Han et al., 2025). Addressing this relationship requires a comprehensive approach that examines both indoor and outdoor environments and considers the broader urban context in which these interactions occur.

Despite the growing recognition of these factors, two key gaps remain in the planning literature. First, studies often fail to holistically examine the interplay between indoor and outdoor settings. While business and marketing research delves into indoor elements, planning literature tends to focus solely on outdoor aspects, neglecting the combined impact of these environments on customer satisfaction.

Second, the influence of the urban context, particularly the dominant mode of transportation in a neighborhood, on place perceptions has been underexplored. Transportation research indicates that people's perceptions of a built environment vary based on their previous travel behaviors, such as their primary mode of travel (Handy, 2005). For instance, in car-centric suburbs, street aesthetics may hold less significance compared to pedestrian-oriented areas where walking significantly shapes the urban experience.

This study addresses these gaps by investigating how place, street, and neighborhood-level characteristics collectively influence customer satisfaction and how these dynamics vary with the level of car dependency in surrounding neighborhoods (Figure 1). Using customer reviews, UGPs, and rating scores sourced from Yelp, this research captures user perceptions and satisfaction levels of EDEs. Additionally, the study utilizes Google Street View images surrounding each EDE to estimate the perceived safety of the neighborhood environment. Computer vision models are used to quantify people's perception to indoor and outdoor images. Ordinal logistic regression is employed to assess the relative significance of these characteristics in shaping customer satisfaction. Moreover, the study considers the diversity of urban contexts, recognizing that insights from pedestrian-friendly areas may not directly apply to car-dependent neighborhoods. This contextual diversity is critical for generalizing the findings and tailoring strategies for different cities.

Advances in AI and computer vision have provided novel tools to analyze the visual and perceptual qualities of a place, thereby transforming the study of urban environments. By integrating insights from marketing, planning, and transportation research, this study bridges disciplinary gaps and provides a comprehensive framework for understanding place satisfaction. The findings offer actionable recommendations for urban planners, designers, and policymakers aiming to enhance the appeal of destinations with EDEs. In addition to reinforcing the link between walkability and vibrant local businesses, this research contributes to broader initiatives in sustainable urban development, public health, and smart city design.

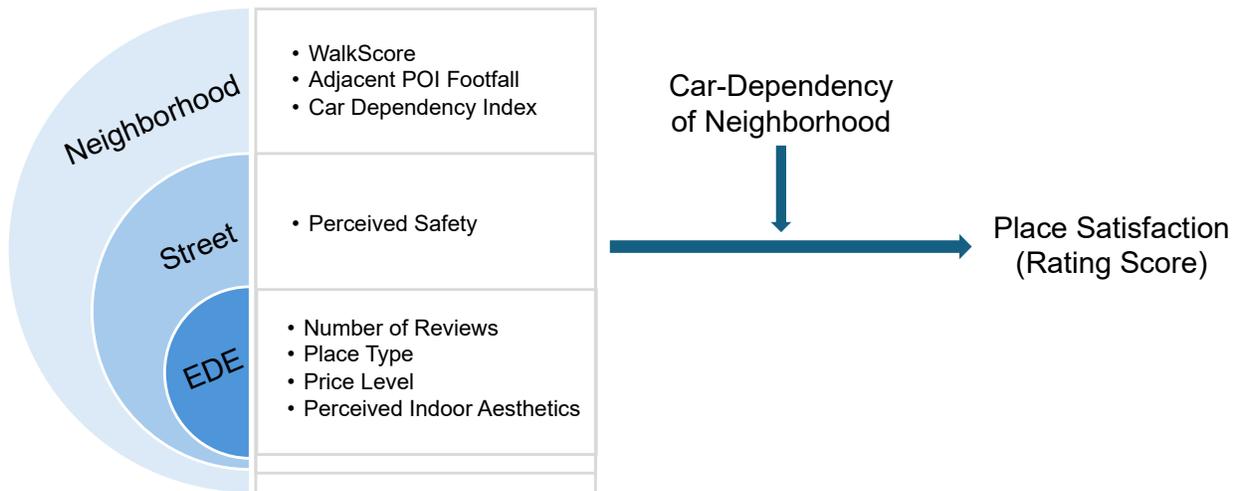

Figure 1: Conceptual Framework of Study

## 2. Literature Review

### 2.1. Place Characteristics and Customer Satisfaction

In the business and marketing literature, the physical environment within EDEs is recognized as a critical factor influencing customer satisfaction and patronage. The physical environment of an EDE includes elements such as furniture/seating arrangement and quality, lighting, color schemes, and architectural style, which play significant roles in shaping the dining experience (Ryu & Jang, 2008; Horng & Hsu, 2021; Lin & Worthley, 2012; Miles et al., 2012). Furthermore, other studies highlight the impact of intangible factors within these indoor settings, such as scent (Chebat & Michon, 2003), temperature (Heung & Gu, 2012), and background music (Morrison et al., 2011), all of which significantly affect customer behavior and satisfaction.

While previous studies mostly examined the factors using surveys, recent advancements in computer vision technologies and the use of crowd-sourced, user-generated data have enabled a broader and more scalable investigation into how the aesthetics of a place influence customer behavior and satisfaction. For example, Zhang and Luo (2023) analyzed restaurant review images sourced from Yelp and applied a computer vision model using the Clarifai API, which classifies food-related photos. Their findings indicated that user-generated images have substantial predictive power in determining the survival of independent, young to middle-aged, and mid-priced restaurants. Similarly, Li et al. (2023) investigated the impact of user-generated photos on hotel choices. They utilized the YOLOv3 object detection model to classify images, such as those of guest rooms and food and beverages. Additionally, they extracted information on the brightness of the photos and integrated it into their model. These studies demonstrate the growing potential of leveraging advanced computer vision techniques and user-generated content to gain deeper insights into the factors that drive customer satisfaction.

## 2.2. Street and Neighborhood Characteristics and Customer Satisfaction

Environmental psychology and planning literature have focused more on the external and contextual characteristics of the surrounding streets and neighborhoods, and their impact on customer satisfaction and destination choices (Kumar et al., 2013). Accessible, safe, comfortable, and aesthetically pleasing streets are theorized to positively influence people's choices of destinations. Numerous studies have sought to quantify these elements to better understand their impact.

Urban form is recognized as a critical macro-scale factor shaping neighborhood and street accessibility and convenience. This includes features such as density, mixed land use, connectivity, and retail floor area ratio (Frank et al., 2006). These characteristics not only make streets more functional and navigable but also contribute to the vibrancy and attractiveness of urban spaces, enhancing their appeal as destination choices.

Beyond urban form, micro-scale street features such as sidewalk quality, street furniture, lighting, and greenery also significantly influence the attractiveness of streets and neighborhoods (Ewing & Handy, 2009; Gehl, 2011). High-quality sidewalks, for example, provide safety and comfort for pedestrians, encouraging more foot traffic and interaction with street-front businesses, and even more public transit use (Park et al., 2021). Street furniture and lighting enhance the functionality and safety of the environment, while greenery and landscaping contribute to the aesthetic appeal, which has been shown to positively affect individuals' perceptions and willingness to spend time in an area (Jacobs, 1961; Nasar, 1994).

Perceived safety and comfort are also key factors in destination choice. The presence of well-lit, clean, and well-maintained streets promotes a sense of security, making them more attractive as destinations (Painter, 1996). Streets with good visibility, minimal traffic conflicts, and pedestrian-friendly infrastructure, such as crosswalks, signals, and buffers are more likely to encourage people to visit, walk, and engage with the surrounding environment (Dumbaugh & Li, 2010, Dumbaugh & Rae, 2009; Nag et al., 2020; Sevtsuk et al., 2021).

The social and cultural vibrancy of streets and neighborhoods also play a critical role in influencing destination choices. Environments that offer diverse social experiences, such as street events, public art installations, and gathering spaces, attract more visitors and contribute to the overall appeal of the area (Whyte, 1980). The integration of public spaces that facilitate social interaction and leisure activities has been shown to enhance the attractiveness of destinations, contributing to their popularity and frequent use (Gehl, 2011; Montgomery, 1998).

The application of advanced technologies, such as geographic information systems (GIS), computer vision, and crowd-sourced data, has expanded the ability to measure and analyze the influence of street and neighborhood characteristics on pedestrian mobility and destination choices (Li et al., 2016; Sevtsuk et al., 2021; Biljecki et al., 2023; Han et al., 2025; Koo et al., 2022; Lieu & Guhathakurta, 2025;). Researchers have increasingly used GIS-based approaches to map and analyze street networks, land use patterns, and accessibility metrics (Moudon et al., 2006). Moreover, the integration of computer vision techniques with street-view imagery enables the assessment of streetscape quality and aesthetic appeal in a scalable manner (Koo et al., 2022; Lieu & Guhathakurta, 2025; Biljecki et al., 2023). For instance, Sevtsuk et al. (2021) combined GPS trajectory data, GIS layers, and Google Street View panoramas to identify street-level features, such as the number of turns, Green View Index, and Sky View Factor, associated with pedestrian preferences in San Francisco. Similarly, Yang et al. (2022) used a deep learning model (VGG-16, trained on the Places365 dataset) to classify images of urban public space along the Haihe River in Tianjin, demonstrating how social media imagery and machine learning can be used to extract visual preferences at scale.

These technological innovations enable a more comprehensive and scalable assessment of how physical and social environments influence destination choices. However, their application also raises concerns around data representativeness and algorithmic bias. User-generated content, such as photos and reviews, often skews toward the preferences of more affluent, educated, and digitally engaged populations (Tasse & Hong, 2017), leaving out marginalized or underserved groups who seldom visit restaurants or use place rating platforms. As a result, models trained on such content may inadvertently hold existing social biases. In urban planning and design contexts, the critical evaluation of these tools is essential to avoid reinforcing inequities and to ensure that analyses reflect diverse perspectives.

In summary, while the indoor environment of EDEs remains important, the street and neighborhood context also play a critical role in shaping destination choices. Yet, these dimensions are rarely examined together in an integrated framework. This study addresses this gap by jointly assessing the perceived qualities of both the indoor environment and the surrounding areas, offering a more holistic model of place satisfaction. Understanding how these interconnected factors shape consumer preferences is essential for promoting vibrant, inclusive, and sustainable urban areas.

3. Methods

3.1 Data

3.1.1 Selection of the Study Area

This study aims to explore how different levels of car dependency influence the relationship between the built environment and user satisfaction with urban destinations. To achieve this, Washington, DC, was selected as an ideal case to study given its balanced distribution of car modal share. Data on transportation modes to work by census block group (CBG) were first collected from the American Community Survey (ACS) across 382 Metropolitan Statistical Areas (MSAs) in the United States (U.S. Census Bureau, 2022a). With an average car modal share of 0.43, Washington, DC, demonstrates a relatively lower reliance on cars compared to the MSA average and presents a bell-shaped distribution (Appendix Figure A1). This research assumes that car dependency is associated with high rates of car travel, characterized by car-oriented land use patterns and limited alternative transportation options (Litman & Laube, 2002). Consequently, the bell-shaped distribution, centered around a 0.43 car modal share, enables a balanced selection of EDEs reflecting a wider range of car dependency levels within a single study area.

3.1.2 Characteristics of EDEs and their Customer Ratings

To collect EDE-specific data on price level, place type, rating score, number of reviews, and interior images, this study employed web-scraping from Yelp. All EDEs across 9 categories available on the website are selected. The categories are divided into two groups: one represents the price levels (low: $, $$; mid: $$$; and high: $$$$), and the other indicates the type of dining place, including restaurants, bars, and cafes. For example, after entering "Washington DC" in the "address, neighborhood" box, "restaurant" in the "things to do" box and the "$", "$$" button on the website are selected to retrieve all restaurants with "low" price levels located in and around the Washington DC area. Each category search returns up to 240 EDEs, which is the maximum number of results that can be retrieved from the website. While some categories have fewer results, the final sample consists of a total of 744 unique EDEs (Table 1). Figure 2 displays the location of EDEs in the study area, with colors representing the type of place.

Table 1.  Number of EDEs per Place Type and Price Level.

| Place type \ Price level | Low ($, $$) | Mid ($$$) | High ($$$$) | Total |
|---|---|---|---|---|
| Restaurant | 121 | 146 | 149 | 416 |
| Bar | 21 | 107 | 35 | 163 |
| Cafe | 42 | 113 | 10 | 165 |
| Total | 184 | 366 | 194 | 744 |



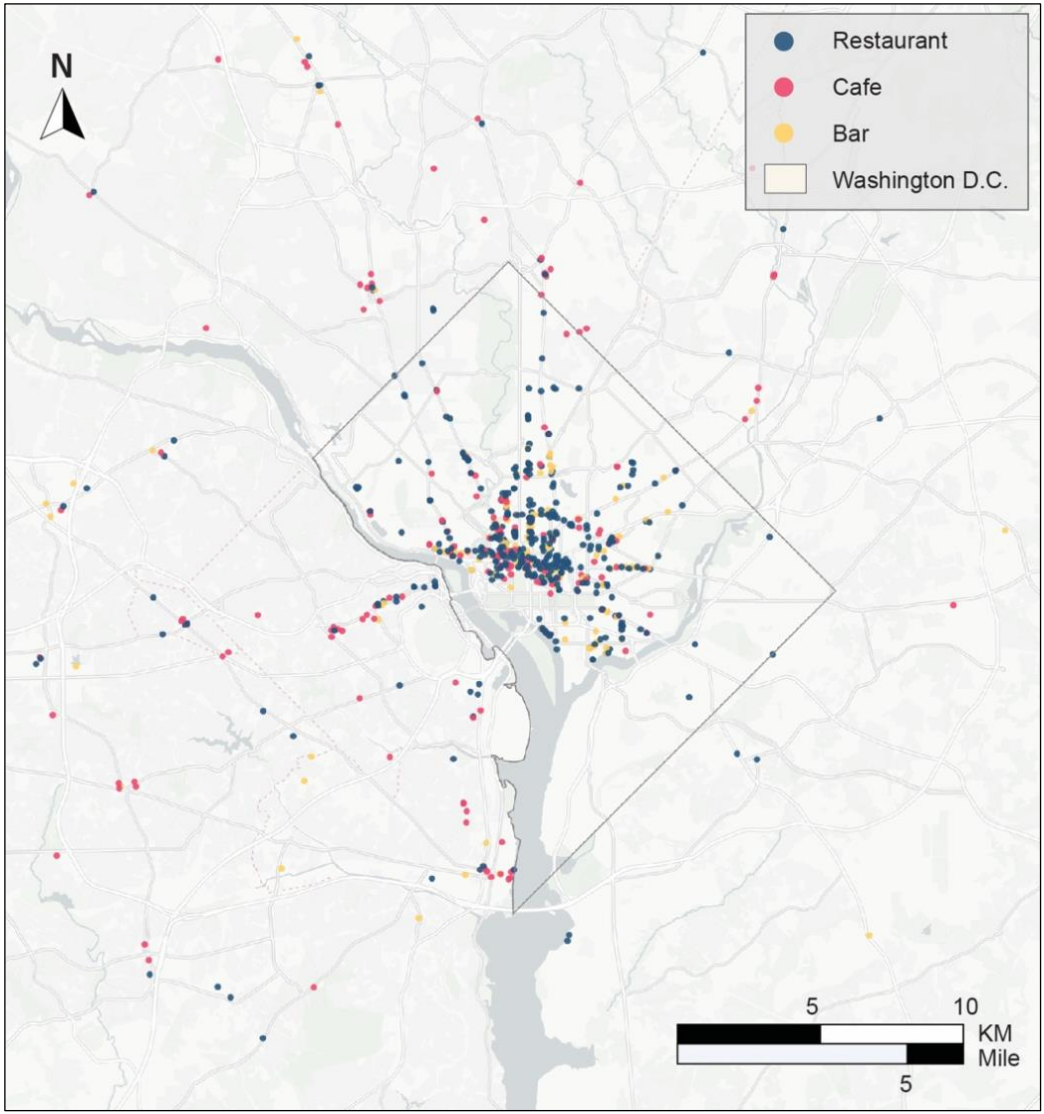

Figure 2. Distribution of EDEs in the study area.

To quantify customer satisfaction and perceptions of indoor space quality, the study utilized rating scores and user-generated photos from Yelp, respectively (Yelp, 2024). The number of reviews was also collected to control for potential biases in customer satisfaction based on review volume when predicting the overall satisfaction level of a place.

For the measurement of interior aesthetics, the study first scraped up to 10 photos categorized as 'indoor' by Yelp from the selected EDEs. After the initial collection, photos that did not accurately represent indoor spaces were manually filtered out. For instance, images that showed only isolated elements, such as a single painting on a wall without any context, or selfies where the indoor environment



was only partially visible, were excluded from further analysis. On average, 8.3 indoor images were collected per EDE.

Then the study assessed the visual appeal of the collected review images using a pre-trained Deep Convolutional Neural Network (DeepCNN) model based on the Aesthetics and Attribute Database (AADB) dataset (Malu et al., 2018; Pan et al., 2024). The AADB dataset, originally created by Kong et al. (2016), consists of manually labeled photos with aesthetic scores and a set of eight visual quality-related attributes, including elements such as Color Harmony, Light, and Vivid Color (Appendix Table A1). Pan et al. (2024) used this dataset to validate their model's performance on restaurant review images. They conducted a pair ranking test with 10 participants, where each participant was presented with two restaurant photos and asked to select the more visually appealing one. The model achieved an accuracy of 0.775, which improved to 0.832 when ambiguous cases (i.e., where the predicted scores of the two photos in a pair differed by less than 0.05) were excluded.

Figure 3 presents a boxplot showing the distribution of aesthetic scores for interior review images across different price levels. The results indicate that higher-priced places generally have higher indoor space scores, while lower-priced places exhibit a more varied distribution of scores. Figure 4 highlights the top 5 and bottom 5 interior photos from mid-priced EDEs, illustrating the range of visual appeal within this category.

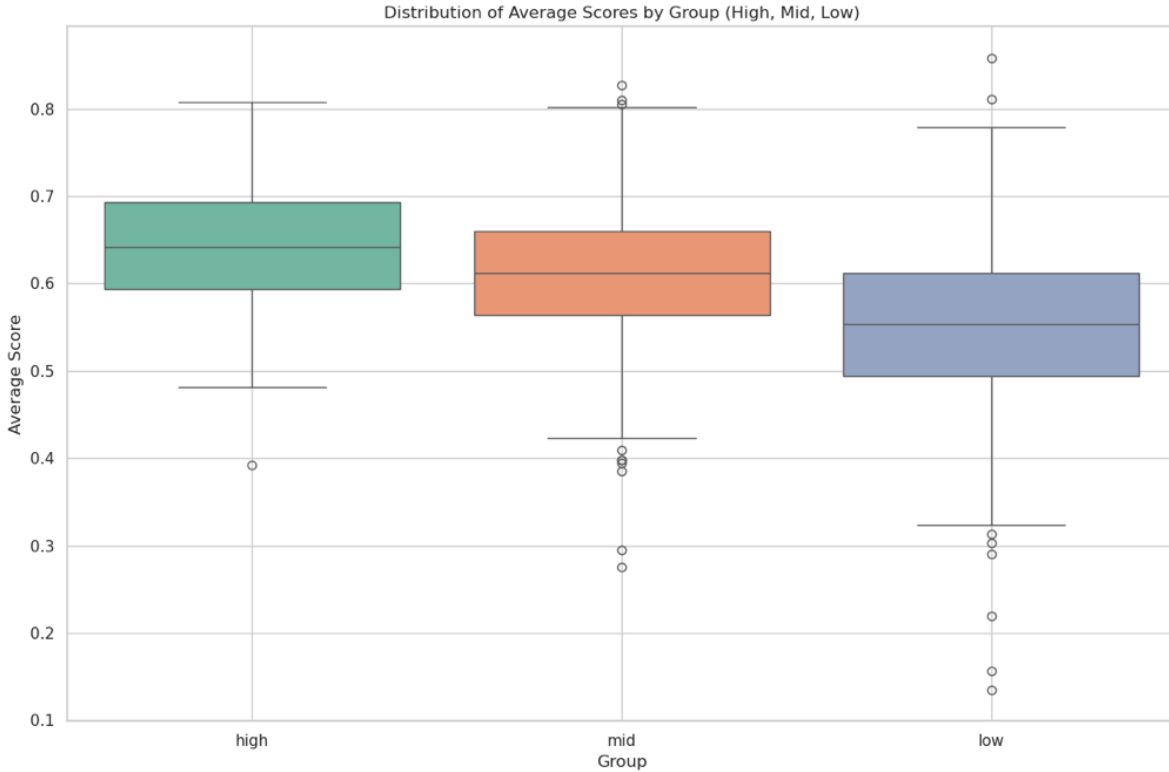

Figure 3. Distribution of Interior Space Aesthetic Score by EDE Type and Price Level



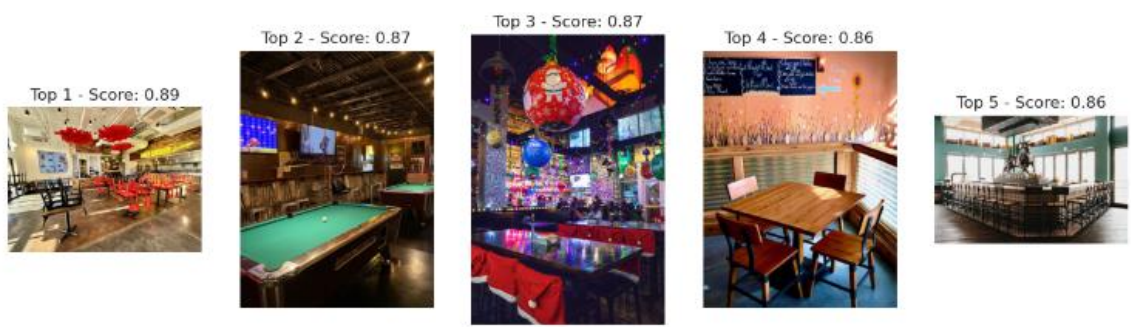

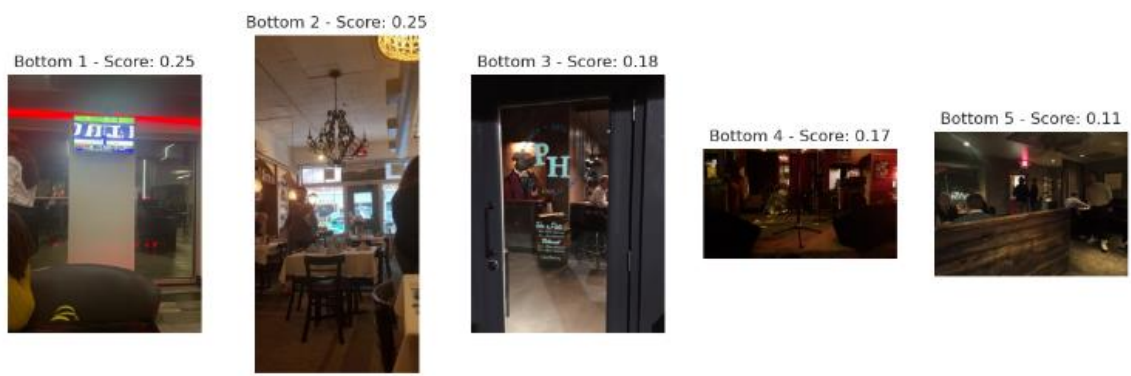

Figure 4. Top and Bottom Five Images of Middle Priced Restaurants in the Study Area.

While the accuracy of the model is in an acceptable range, it is important to note that Yelp's user base may not fully represent the general population (Choi, 2021); for example, 56% of users report higher income and 62% have attended college (Yelp Statistics, 2025). As a result, the content may reflect the preferences of more digitally active or privileged users, potentially underrepresenting other groups. Nevertheless, since our application of computer vision is limited to identifying objective visual features, such as lighting, color balance, and composition rather than subjective evaluations of individuals or behavior, the implications of representational bias are less severe in this context. This method offers an interpretable way to quantify aesthetic features across a large sample of establishments, supporting the broader aim of understanding how interior aesthetics may relate to customer satisfaction.

To account for operational and socio-demographic influences on customer satisfaction, two variables were included in the model: weekly opening hours and the average income of visitors at each EDE. The variable for opening hours is measured as the total weekly hours an EDE operates. This



variable is included to capture potential differences in service quality associated with operational schedules. Places with shorter opening hours often prioritize high-quality service and carefully prepared food, as their limited hours allow more time for staff preparation and resource management (Shy & Stenbacka, 2006). In contrast, venues with extended or 24-hour availability, such as franchise restaurants, often face increased pressure on staff and operational resources. This extended availability may lead to compromised service quality, as these establishments must accommodate continuous demand, often with fewer opportunities for preparation and quality control (Choi et al., 2009). By controlling weekly opening hours, the study aimed to isolate the impact of operational factors that may influence the perception of service and food quality.

The average income of visitors to each EDE was incorporated as dining places tend to attract a relatively homogeneous visitor base by income. Visitor income data for each EDE is calculated using the "Monthly Patterns - Foot Traffic" dataset from Advan Research (2022), which provides aggregated mobile device data on visits to EDEs, linked to the home census tracts of visitors. By combining this dataset with income data from the ACS, the average visitor income for each EDE was calculated, weighing the number of visitors from each home census tract.

$$Average\ Visitor\ Income\ for\ EDE_i = \frac{\sum_j(Visitors_{ij} \times Income_j)}{\sum_j Visitors_{ij}}$$

Where $i$ represents a specific EDE and $j$ represents a specific home census tract, $Visitors_{ij}$ Indicate the number of visitors from home census tract $j$ to EDE $i$ , and $Income_j$ is the median income of census tract $j$. This variable allows us to account for the potential influence of visitors' income on satisfaction ratings, as higher-income visitors may have different expectations and satisfaction levels compared to lower-income visitors.

### 3.1.3 Street Level Characteristics

To measure perceived streetscape safety near EDEs, yet another computer vision-based method was utilized. First, street view images are collected using the Google Street View Static API. At the midpoints of road segments within a 300-meter buffer around each of the 744 EDEs in Washington, DC, two images aligned with road direction but facing opposite directions are captured, resulting in 36,162 total images. Then, a deep learning model developed in Hwang et al. (2023) was applied to quantify visual safety and integrate these scores into our analysis. The model was trained using the Place Pulse 2.0 dataset, developed by MIT Media Lab (Dubey et al., 2016), which contains 1.6 million voting results for 110,988 images from 56 cities. This dataset asked online users to choose which of two street images appeared safer. Using these votes, a model was trained to predict TrueSkill scores, a rating algorithm developed by Microsoft (Herbrich et al., 2006). The model, based on the MaxVit: Multi-Axis Vision Transformer



architecture, was pre-trained on millions of ImageNet images and fine-tuned for the visual safety task using transfer learning. This approach reduced the need for large amounts of new data. Hwang et al. (2023) further refined the model with five additional hidden layers to improve its ability to assess safety cues from urban streetscapes. The model was implemented in PyTorch, utilizing pre-trained weights from the 'timm' library, and integrated the safety scores into the final analysis. While the model enables systematic assessment of perceived streetscape safety at scale, it is trained on a crowdsourced dataset that may reflect certain cultural or demographic perspectives more strongly than others. Thus, the resulting safety scores should be understood as approximations of perception rather than objective measures. Further discussion of potential biases and limitations is provided in the Discussion section.

### 3.1.4 Neighborhood Level Characteristics

WalkScore is calculated based on the distance to 13 categories of amenities, with each category weighted equally (WalkScore, n.d.). The points are then summed and normalized to produce a score ranging from 0 to 100. Despite its simplicity, this metric has been widely used as a reliable tool for estimating access to nearby facilities, which are an important component of accessibility of the built environment (Carr et al., 2010; Hall & Ram, 2018). WalkScores for each EDE location is obtained using the WalkScore API. The mean WalkScore value across EDEs in the study area is 92.4.

To estimate the car dependency of EDEs, representing the extent to which access to the location is reliant on automobiles, the study developed an index incorporating three quantitative measures: car modal share, population density, and employment density. For each EDE, adjacent CBGs whose centroids fall within a 1-km buffer of the EDE were identified. The average car modal share in commuting, as well as the average population and employment density across these adjacent CBGs, was then calculated. Population density was calculated by dividing the total population of each CBG by its area, using data from the ACS 5-Year Estimate (U.S. Census Bureau, 2022a). Similarly, employment density was determined by dividing the total number of jobs in each CBG by its area, using data from the Longitudinal Employer-Household Dynamics (LEHD) Origin-Destination Employment Statistics data (U.S. Census Bureau, 2022b).

Each of the three measures was then normalized using the min-max method to create comparable scales. Based on the normalized values, the research created an overall car dependency index for each EDE using the following equation:

$$CD_i = w \times MS_i + w \times (1 - PED_i)$$

where $CD_i$ represents the car dependency index score of EDE $i$, $MS_i$ denotes the normalized average car modal share of the adjacent block groups, and $PED_i$ is sum of the normalized values for average



population density and employment density of adjacent block groups. To reflect an equal influence of each factor, the study set the weight $w$ for each term to 0.5.

By subtracting $PED_i$ from one, the study account for the inverse relationship: as population and employment density increase, the likelihood of car dependency tends to decrease due to the availability of alternative transportation options in high-density areas. Thus, this index indicates EDEs with a higher car modal share and lower population, and employment densities are more likely to be accessed by car. For easier interpretation of model parameters, the study scaled this variable by multiplying 100, so values range from 0 to 100. On average, EDEs had approximately 14 census block groups with centroids within the 1-km buffer. However, this number varied depending on the size of the census block group, as block group sizes are determined based on population density. Typically, more urbanized areas have smaller census block groups, while suburban or rural areas feature larger ones.

Additionally, the research introduced an interaction term between the calculated car dependency index and the perceived safety of surrounding streets described in the street-level characteristics section. This interaction term allows us to explore how the dependency on cars for accessing EDEs influences the significance of street quality in shaping satisfaction ratings. Previous research in transportation studies suggests that dominant travel modes significantly impact individuals' perceptions of their environment (Gehl, 2010; Handy, 2005; Ewing & Handy, 2009). Accordingly, for visitors who primarily access EDEs by car, the quality of surrounding streets may play a less critical role in their overall satisfaction.

To control neighborhood popularity, the study measured the footfall of visitors to other EDEs within a 500-meter buffer from each sampled EDE. The data on visitor volume of each EDE was sourced from the Monthly Patterns - Foot Traffic dataset provided by Advan Research (2022), providing insights into visit patterns. This measure helps capture the relative popularity and activity density surrounding each EDE, allowing for a more accurate analysis of how neighborhood dynamics influence customer satisfaction.

### 3.2 Ordinal Logistic Regression Model

To investigate the factors affecting customer satisfaction, represented by rating scores, an ordinal logistic regression model is employed. Rating score, our dependent variable, is treated as an ordinal measure divided into five categories: 1.5 - 3.0, 3.0 - 3.5, 3.5 - 4.0, 4.0 - 4.5, and 4.5 - 5.0, with each range including the upper bound. The rating score was regressed on a range of explanatory variables, including price level, place type, interior aesthetic scores, and additional neighborhood and street-level characteristics detailed in Table 2. The ordinal logistic regression model estimates the conditional probability that customer satisfaction (i.e., rating score), in response to the explanatory variables, will be less than or equal to a specific level given the predictor values. Letting $Y$ represent the rating score and $k$



= 5 (i.e., the number of ordered categories), the probability of a rating score $Y$ falling into category $j$ or below is expressed by:

$$P(Y \leq j) = \frac{\exp(\alpha_i + X\beta)}{1 + \exp(\alpha_i + X\beta)}, \quad j = 1, 2, \ldots, k - 1$$

where $\alpha_i$ represents the threshold for each cumulative probability, $X$ is the vector of explanatory variables and $\beta$ is the vector of coefficients for $X$ which is assumed to be constant across all categories $j$, reflecting the proportional odds assumption.

Before estimating model parameters, two key assumptions of ordinal logistic regression were tested: the proportional odds assumption and the absence of multicollinearity. The proportional odds assumption, which posits that odds ratios remain consistent across all thresholds of the ordinal outcome, was assessed using the Brant test. The test results were not significant, indicating that the regression slopes do not differ substantially across levels of the dependent variable, thus supporting the proportional odds assumption. A graphical assessment of the parallel slopes further confirmed the stability of coefficients across outcome levels for each predictor. To evaluate multicollinearity, correlations among the explanatory variables were examined. The correlation matrix did not reveal any strong linear relationships, suggesting minimal risk of multicollinearity. Furthermore, variance inflation factor (VIF) values for each variable ranged from 1.1 to 2.5 - well below the threshold of 5 - confirming that multicollinearity is not a concern (Appendix Table A2).

Table 2. Data Description.

| Variable(s) | | | Mean (Number) | Std. dev. (%) | Min. | Max. |
|---|---|---|---|---|---|---|
| *Dependent Variable* | | | | | | |
| Rating Score | 1.5 - 3.0 | | 28 | 3.8 % | | |
| | 3.0 - 3.5 | | 96 | 12.9 % | | |
| | 3.5 - 4.0 | | 269 | 36.2 % | | |
| | 4.0 - 4.5 | | 306 | 41.1 % | | |
| | 4.5 - 5.0 | | 45 | 6.0 % | | |
| *Independent Variables* | | | | | | |
| Neighborhood Level | Accessibility (WalkScore) | | 92.3 | 11.0 | 15.0 | 100.0 |
| | Adjacent POI Footfall | | 957.3 | 864.1 | 88.4 | 4116.9 |
| | Car Dependency Index | | 54.6 | 20.7 | 1.1 | 99.1 |
| | Neighborhood Income level (10k) | | 10.8 | 2.6 | 1.4 | 23.3 |
| Street Level | Perceived Safety | | 6.0 | 0.3 | 4.7 | 6.6 |
| Point-of-Interest Level | Perceived Aesthetic | | 6.0 | 0.9 | 1.6 | 8.0 |
| | Number of Reviews | | 387.0 | 399.2 | 1.0 | 2369.0 |
| | Opening hours in a week | | 63.7 | 23.7 | 7.0 | 168.0 |
| | Place type | Restaurant | 416 | 55.9 % | | |
| | | Bar | 163 | 21.9 % | | |
| | | Café | 165 | 22.2 % | | |
| | Price level | Low | 184 | 24.7 % | | |
| | | Mid | 366 | 49.2 % | | |
| | | High | 194 | 26.1 % | | |
| | Income Level of EDE Visitors | | 11.5 | 3.8 | 3.1 | 25.0 |

## 4.   Results

Table 3 presents the outcomes of the regression model. Odds ratios, calculated by exponentiating coefficients, indicate the likelihood of higher category rating with changes in the independent variables (Table 4). Also, marginal effects (or marginal percentages), which makes the interpretation of ordinal logistic regression results easier, were calculated. Marginal effects represent the change in the probability of each outcome category for a one-unit change in a predictor variable, holding other variables constant. The results are summarized into three categories: neighborhood, street, and POI-level characteristics.

Table 3. Ordinal Logistic Regression Model Results

| Variable | | Coef. | | Std. error | p-value |
|---|---|---|---|---|---|
| Neighborhood Level | Accessibility (WalkScore) | 0.030 | *** | 0.008 | 0.000 |
| | Adjacent POI footfall | - 0.001 | *** | 0.000 | 0.000 |
| | Income Level of Neighborhood | 0.005 | | 0.029 | 0.874 |
| | Car Dependency Index (CDI) | 0.432 | *** | 0.023 | 0.000 |
| Street Level | Perceived Safety | 4.468 | *** | 0.166 | 0.000 |
| | Perceived Safety × CDI | - 0.068 | *** | 0.004 | 0.000 |
| Point-of-Interest level | Income Level of EDE Visitors | 0.030 | | 0.041 | 0.467 |
| | Perceived Aesthetic | 0.307 | *** | 0.090 | 0.000 |
| | Number of Reviews | 0.000 | * | 0.000 | 0.013 |
| | Opening Hours in a Week | - 0.020 | *** | 0.003 | 0.000 |
| | Place Type (ref: bar)   Cafe | 0.890 | *** | 0.224 | 0.000 |
| | Restaurant | 1.774 | *** | 0.196 | 0.000 |
| | Price Level   Linear | - 1.057 | ** | 0.392 | 0.007 |
| | Quadric | - 0.332 | | 0.352 | 0.346 |
| | Opening Hours × Price Level (Linear) | 0.012 | * | 0.006 | 0.035 |
| Number of observations | | 744 | | | |
| McFadden's R-squared (d.f. = 20) | | 0.112 | | | |

| AIC | 1746.283 |
|---|---|

Table 4. Odds Ratios of the Ordered Logistic Model

| Variable | | | Odds Ratio |
|---|---|---|---|
| Neighborhood Level | Accessibility (WalkScore) | | 1.03 |
| | Adjacent POI footfall | | 1.00 |
| | Car Dependency Index (CDI) | | 1.54 |
| Street Level | Perceived Safety | | 87.18 |
| | Perceived Safety × CDI | | 0.93 |
| Point-of-Interest level | Perceived Aesthetic | | 1.36 |
| | Number of Reviews | | 1.00 |
| | Opening Hours in a Week | | 0.98 |
| | Place Type (ref: bar) | Cafe | 2.44 |
| | | Restaurant | 5.89 |
| | Price Level | Linear | 0.35 |
| | Opening Hours × Price Level (Linear) | | 1.01 |

**Neighborhood-level effects**

*WalkScore:* The WalkScore variable shows a positive association with rating score, with a coefficient of 0.03. This indicates that for each 1-point increase in the WalkScore, the likelihood of a place having a higher rating score category would be slightly higher by 3% ($\exp(0.03) \approx 1.03$), holding other factors constant. This suggests that EDEs located in more walkable areas, with better access to amenities and services, tend to receive higher customer satisfaction ratings.

*Adjacent POI:* The adjacent POI footfall has a coefficient of -0.001, indicating a negative and statistically significant relationship with rating score. The odds ratio ($\exp(-0.001) \approx 0.999$) suggests that a 1-unit increase in neighborhood popularity (measured as footfall) slightly decreases the likelihood of a higher rating score category by about 0.1%. This negative association may reflect that EDEs in high-footfall areas can experience greater congestion, potentially impacting customer satisfaction ratings. Interestingly, this finding contradicts Whyte's (1980) observation that people are drawn to places where others are present. While some levels of surrounding activity may be attractive, excessive intensity of adjacent POIs and foot traffic may lead to overcrowding, diminishing the overall experience. This result suggests that the relationship between surrounding activity and perceived quality may follow a non-linear pattern, where moderate vibrancy is appreciated, but overcrowding can be unappealing.

*Median income of neighboring areas:* The income level of neighboring census block groups did not show statistically significant relationship with rating score.

*Car dependency:* The Car Dependency Index (CDI) has a positive relationship with customer satisfaction, with a coefficient, indicating that higher car dependency is associated with higher rating scores, when all others hold constant. It is possibly because these locations may offer convenient automobile access, which aligns with automobile-oriented consumers' preferences. Particularly for suburban or rural areas where driving is the primary mode of transportation, and customers value ease of access, this result is expected.

**Street-level effects**

*Perceived safety:* The Perceived Safety Score, derived from Google Street View images around each EDE, was found to have a positive and statistically significant relationship with the rating score. The odds ratio of 87.18 indicates that for every 0.1-point increase in the perceived safety score (ranging from 0 to 1), the likelihood of rating score being in a higher category would increase by 8.7 times, holding all other variables constant. This suggests that higher levels of perceived safety around an EDE are associated with higher customer satisfaction ratings.

However, the interaction term between perceived safety score and CDI shows a negative coefficient, suggesting that the effect of perceived safety score on customer satisfaction diminishes when car dependency is higher. Also, the negative interaction term of CDI with the perceived safety score highlights that when an area is perceived as safe, the advantage of car accessibility diminishes. In other words, in safer neighborhoods, people may prefer walking or other modes of transportation, and the reliance on cars becomes less of a determinant for customer satisfaction.

**EDE related factors and their effects**

*Perceived aesthetic score:* The perceived aesthetics score, derived from the Yelp review images, is significantly and positively associated with the rating score, indicating that aesthetically pleasing EDEs are strongly associated with higher satisfaction. The odds ratio of 1.36 indicates that with a 1-point increase in perceived aesthetics score ranging from 0 to 10, the likelihood of a place having a higher category rating score would be 1.36 times greater.

*Number of reviews:* While the number of reviews appears to have a statistically significant relationship with rating score, the coefficient was miniscule. Also, the income level of EDE visitors did not show statistical significance in the model.

*Type and price level of EDEs:* The type and price level of an EDE were also significant factors influencing customer satisfaction. Compared to bars, cafes and restaurants had higher rating scores, with positive odds ratio of $\exp(0.89) \approx 2.43$ and $\exp(1.77) \approx 5.87$, respectively. This suggests that customers tend to rate cafes and restaurants more favorably than bars. Additionally, price level showed a significant negative and linear relationship with rating score, indicating that people would be more satisfied with cheaper options, when other variables are holding constant.

*Opening hours:* Opening hours per week had a slight negative relationship with the rating score. A one-hour increase in weekly opening hours is associated with a 2% decrease in the likelihood of a place having a higher category rating score. This may suggest that extended operating hours, which can lead to greater wear on facilities and potentially strained staff (e.g., 24-hour restaurants), could have a modestly negative impact on customer satisfaction. The positive coefficient found in the interaction term between opening hours and price level indicates that as either price level or opening hours increase, the effect of the other variable becomes stronger.

## 5. Discussion

This study demonstrates the critical role of both indoor and outdoor environments in shaping customer satisfaction with dining destinations. The strong association between interior aesthetics and customer ratings aligns with the servicescape literature, which emphasizes the impact of ambiance on customer experience (Bitner, 1992; Ryu & Jang, 2008). However, this study extends these insights by integrating exterior factors, such as perceived street safety, and car-dependency, illustrating the relationships between indoor and outdoor elements in determining satisfaction in urban settings with different levels of car dependency.

The positive association between perceived safety and place satisfaction reflects findings in walkability research, where safe and accessible streets are linked to greater user engagement and satisfaction (Jacobs, 1961; Ewing & Handy, 2009; Koo et al. 2021). Yet, the moderating effect of car dependency suggests the need for a context-sensitive approach. In car-dependent areas, the convenience of driving overshadows the benefits of safe streetscapes. This finding aligns with transportation research that highlights how dominant travel modes shape environmental perceptions (Gehl, 2011; Handy, 2005; Ewing & Handy, 2009).

The finding suggests that even in places where driving is the dominant mode of transportation, streetscape quality remains an important factor influencing customer satisfaction. Therefore, rather than viewing interior enhancements and external streetscape improvements as competing priorities, these elements should be seen as complementary. In car-dependent areas, enhancing multimodal accessibility, such as integrating pedestrian, cycling, and transit infrastructure, can simultaneously elevate streetscape quality and address customer needs, achieving both immediate satisfaction and long-term sustainability goals. Rather than expanding parking facilities or designing car-centric layouts, planners and businesses can prioritize strategies that improve the pedestrian experience while accommodating existing car users. For instance, strategically located parking hubs connected to pedestrian pathways or transit stops can encourage walking and transit use without sacrificing convenience for drivers. Investments in well-lit walkways, safe crossings, and accessible streetscape features can create environments that appeal to pedestrians and cyclists, thereby fostering vibrant and inclusive urban spaces.

Businesses in such areas can play an active role by designing interiors and storefronts that engage with surrounding spaces, creating a more inviting and walkable environment. Investing in green infrastructure, outdoor seating, and aesthetic enhancements that appeal to pedestrians can create a seamless connection between the interior environment and the surrounding urban context. These strategies not only address customer satisfaction but also align with long-term sustainability goals by fostering a shift away from car dependence.

In walkable neighborhoods, where pedestrian traffic is already significant, enhancing streetscape aesthetics can yield higher satisfaction and economic vitality. Investments in measures such as adding greenery, reducing traffic speeds, and creating pedestrian-friendly public spaces can further strengthen the appeal of these areas. Improving pedestrian safety through well-marked crosswalks and incorporating amenities like shaded walkways or public seating areas can encourage more foot traffic and bolster local businesses.

By emphasizing sustainability, this approach aligns with broader urban development goals, such as reducing carbon emissions, promoting active transportation, and enhancing public health. Walkable environments not only contribute to vibrant commercial districts but also foster inclusive and livable urban spaces. Recognizing the diverse needs of different contexts, this study provides suggests targeted interventions that balance immediate accessibility with long-term sustainability.

There are several limitations to this study. First, the Car Dependency Index, while useful, may not fully capture whether visitors actually use cars to access EDEs. While a higher index score indicates that an EDE is located in an area with greater car reliance, and a lower score suggests lower reliance, this index does not account for the modal share or trip distance from the trip's origin. Since the index is based solely on the area surrounding the EDE, it fails to consider trip-level factors, such as the car modal share of the origin area and the distance between the origin and the POI (i.e., the destination). For example, despite a higher car dependency index score in this study, individuals might still prefer walking for shorter trips. Conversely, individuals may choose to drive to an EDE because they reside in an area with a higher car modal share, even if the EDE has a lower car dependency score. These exceptions can distort the accuracy of the index at the trip level. Therefore, more sophisticated methods for measuring car dependency need to be developed to better represent the actual car dependency of EDEs.

Second, there may be a bias in measuring the perceived safety of streetscapes around each EDE due to the image coverage area. This study collected street view images at the midpoints of road segments within a 300-meter buffer around each EDE. However, this 300-meter buffer may not fully represent the average streetscape due to the variance in the number of images captured. The average number of images per EDE is 48.5, with a standard deviation of 17.2, a minimum of 2, and a maximum of 106, indicating that some EDEs have significantly more images than others. This discrepancy may fail to capture the streetscape as perceived by individuals, as people may not be influenced by road segments located far from the EDE. In future studies, measuring only the streetscape of the closest street to the EDE could provide a more accurate assessment.

Also, as mentioned earlier, the assessment of perceived streetscape safety relies on a computer vision model trained on a crowd-sourced dataset, which aggregates subjective safety judgments from a global pool of online participants. This approach inherently introduces potential biases rooted in the cultural, racial, and socioeconomic backgrounds of the raters. Prior research has shown that perceptions of urban safety can vary significantly based on individuals' gender and lived environments (e.g., country of residence), suggesting that computer vision models trained on crowdsourced data may capture only a subset of these diverse perspectives (Cui et al., 2023; Quintana et al., 2025).

Additionally, while our model demonstrates strong predictive performance, it remains opaque ("black box") in how it weighs different visual features. Prior studies suggest that elements such as visible greenery, good lighting, wide sidewalks, and the absence of litter or graffiti often correlate with higher perceived safety (Li et al., 2015; Yu et al., 2025). Future work could apply explainable AI techniques (e.g., saliency maps) to better understand the specific features that contribute to safety predictions.

Finally, the assessment of interior aesthetics using user-generated photos may not fully capture the preferences of the broader population. As noted in Section 3.1.2, the aesthetic model was trained on the AADB dataset, which lacks demographic information about photo reviewers. Similarly, the validation study by Pan et al. (2024) does not report participant demographics, limiting our ability to evaluate whether the model reflects a representative range of aesthetic preferences. While the model focuses on objective visual features (e.g., symmetry, color harmony), underlying biases in aesthetic values may still exist. Future research should investigate these potential biases and explore ways to incorporate more representative or culturally diverse aesthetic datasets.

By addressing these limitations and building on the insights presented in this paper, future researchers can deepen our understanding of the relationships between built environments, its urban context, and customer satisfaction. These findings highlight the importance of aligning short-term interventions with long-term sustainability goals, providing actionable strategies to enhance urban environments, foster vibrant local economies, and create more sustainable and inclusive cities.

6.  Conclusion

In conclusion, this study contributes to the existing literature by examining how customer satisfaction with EDEs is influenced by both interior and exterior built environment characteristics across different urban contexts, particularly focusing on the role of car dependency. By integrating data from Yelp reviews and Google Street View images, the study quantified the impact of various servicescape factors, including interior aesthetics, perceived safety, and streetscape walkability, on customer ratings. The findings reveal that while the interior environment consistently plays a significant role in shaping customer satisfaction, the influence of exterior factors such as perceived safety varies depending on the car dependency level of the area.

In car-dependent areas, the advantage of perceived safety and walkable streetscapes diminishes, as customers prioritize accessibility and the convenience of driving over the quality of the walking environment. In contrast, in walkable neighborhoods, streetscape and perceived safety become crucial factors that enhance customer satisfaction. These insights highlight the need for a context-sensitive approach in urban planning, where strategies to boost customer satisfaction and local business success are tailored based on the transportation characteristics and built environment of the area. By doing so, urban planners and designers can more effectively align interventions, such as enhancing interior spaces, improving streetscapes, or improving multi-modal accessibility, according to the specific needs of different city types, ultimately fostering more vibrant and successful commercial districts. This study emphasizes the critical importance of understanding the interaction between transportation modes and environmental aesthetics in designing cities that support both business vitality and urban livability.

Finally, this research underscores the growing potential of user-generated data and computer vision techniques for planning and design applications. Yelp reviews and Google Street View imagery represent scalable and cost-effective tools for evaluating the physical and perceptual qualities of urban environments. These data sources can complement traditional site audits and survey-based assessments by offering insights into how everyday users perceive and interact with places. However, limitations such as sampling bias and the interpretability of computer vision outputs must be acknowledged. Moving forward, integrating these digital tools more systematically into planning workflows can help cities better assess, design, and manage environments that are responsive to both aesthetic and functional needs of diverse urban populations.

**Acknowledgements**

This research did not receive any grant from funding agencies in the public, commercial, or not-for-profit sectors.

**Declaration of Interest**

The authors declare that they have no known competing financial interests or personal relationships that could have appeared to influence the work reported in this paper.

# Appendix

https://github.com/bravoyourlif/aestheticpoi/blob/main/JUD_Appendix.pdf